\documentclass[%
 reprint,
 twocolumn,
 superscriptaddress,
 amsmath,amssymb,
 aps,
]{revtex4-2}

\usepackage{graphicx}% Include figure files
\usepackage{dcolumn}% Align table columns on decimal point
\usepackage{bm}% bold math
\usepackage{color} 
\begin{document}

%\preprint{APS/123-QED}

\title{Copula-based analysis of the generalized friendship paradox in clustered networks}% 
%\title{Vine copula method for analyzing the generalized friendship paradox}% 

\author{Hang-Hyun Jo}
\affiliation{%
 Department of Physics, The Catholic University of Korea, Bucheon 14662, Republic of Korea
}%

\author{Eun Lee}
\affiliation{Department of Scientific Computing, Pukyong National University, Busan 48513, Republic of Korea}

\author{Young-Ho Eom}
\email{yheom@uos.ac.kr}
\affiliation{Department of Physics, University of Seoul, Seoul 02504, Republic of Korea}
\affiliation{Urban Big data and AI Institute, University of Seoul, Seoul 02504, Republic of Korea}

\date{\today}% It is always \today, today,
             %  but any date may be explicitly specified
             
\begin{abstract}
A heterogeneous structure of social networks induces various intriguing phenomena. One of them is the friendship paradox, which states that on average your friends have more friends than you do. Its generalization, called the generalized friendship paradox (GFP), states that on average your friends have higher attributes than yours. Despite successful demonstrations of the GFP by empirical analyses and numerical simulations, analytical, rigorous understanding of the GFP has been largely unexplored. Recently, an analytical solution for the probability that the GFP holds for an individual in a network with correlated attributes was obtained using the copula method but by assuming a locally tree structure of the underlying network [Jo~et~al., Physical Review E~\textbf{104}, 054301 (2021)]. Considering the abundant triangles in most social networks, we employ a vine copula method to incorporate the attribute correlation structure between neighbors of a focal individual in addition to the correlation between the focal individual and its neighbors. Our analytical approach helps us rigorously understand the GFP in more general networks such as clustered networks and other related interesting phenomena in social networks. 
\end{abstract}

\maketitle

\section{Introduction}\label{sec:intro}

Complex systems have been characterized in terms of graphs or networks~\cite{Albert2002Statistical, Borgatti2009Network, Barabasi2016Network, Newman2018Networks, Menczer2020First}: Nodes of the network represent elements of the system, while pairwise interactions between elements are denoted by links between nodes. Thanks to the rapid development of information-communication technology researchers have access to large-scale digital records for human social behaviors, enabling us to study the social networks in more detail. By the empirical analyses of a variety of social network datasets it has been shown that social network topology is heterogeneous in many aspects~\cite{Jo2018Stylized}, often characterized by heavy-tailed distributions of degrees~\cite{Barabasi1999Emergence, Broido2019Scalefree}, assortative mixing or homophily~\cite{Newman2002Assortative}, and community structure~\cite{Fortunato2010Community} to name a few. Such a heterogeneous structure of social networks induces a number of intriguing phenomena in complex social networks. They include dynamical processes taking place on those social networks, such as spreading and diffusion~\cite{Castellano2009Statistical, Sen2014Sociophysics, Pastor-Satorras2015Epidemic, Masuda2017Random}.

One of the interesting phenomena induced by the heterogeneous structure of social networks is the friendship paradox (FP), which states that on average your friends have more friends than you do~\cite{Feld1991Why}. The FP focuses on the number of neighbors of the node, i.e., the node degree. However, nodal attributes other than the degree can also be used for the comparison of an individual to its neighbors, leading to the generalized friendship paradox (GFP)~\cite{Eom2014Generalized}. Candidates for nodal attributes could be either topological, e.g., betweenness centrality~\cite{Grund2014Why} and eigenvector centrality~\cite{Grund2014Why, Higham2019Centralityfriendship}, or non-topological, e.g., scientific achievement~\cite{Eom2014Generalized}, happiness~\cite{Bollen2017Happiness}, and sentiment~\cite{Zhou2020Sentiment}. The GFP states that on average your friends have higher attributes than yours~\cite{Hodas2013Friendship, Eom2014Generalized, Jo2014Generalized}. The GFP has been extensively investigated not only by analyzing various empirical datasets~\cite{Hodas2013Friendship, Eom2014Generalized, Lerman2016Majority, Momeni2016Qualities, Benevenuto2016Hindex, Bollen2017Happiness, Alipourfard2020Friendship, Zhou2020Sentiment, PloegerMansueli2022Friendship} but also by means of numerical and analytical approaches~\cite{Jo2014Generalized, Fotouhi2015Generalized, Higham2019Centralityfriendship, Jo2021Analytical, Cantwell2021Friendship}.

Both FP and GFP have been studied at the network level as well as at the individual level. A typical approach at the network level is to compare the expected degree or attribute of a randomly chosen node to that of an end node of a randomly chosen link~\cite{Feld1991Why, Eom2014Generalized}. On the other hand, the individual-level approach has been studied in terms of the probability that the degree or attribute of a focal node in a network is smaller than the average degree or attribute of neighbors of the focal node. This probability can be interpreted as peer pressure~\cite{Lee2019Impact}. From now on, nodal attributes indicate only non-topological ones to distinguish from degrees for the FP. Focusing on the GFP, nodes with high degree and high attribute are generically expected to have lower peer pressure than other nodes. Interestingly, it was also found by empirical analysis of scientific collaboration networks~\cite{Eom2014Generalized} that some nodes with high degree and high attribute have higher peer pressure in a network with positively correlated attributes, i.e., a homophilic network~\cite{McPherson2001Birds}, than expected for random counterparts. It is because when attributes of neighboring nodes are positively correlated with each other, nodes with high degree and high attribute are likely to be surrounded by neighbors with even higher degree and higher attribute than themselves. This empirical finding was supported by numerical simulations of a minimal model with tunable correlations between attributes of neighboring nodes~\cite{Jo2014Generalized}.

Although the analytical solution for the GFP in a network with uncorrelated attributes was obtained several years ago~\cite{Jo2014Generalized}, analytical, rigorous understanding of the GFP in a network with correlated attributes has been largely unexplored due to the lack of proper mathematical tools for modeling the correlation structure between attributes of neighboring nodes. Recently, Jo et al.~\cite{Jo2021Analytical} modeled the correlation structure between attributes of neighboring nodes by means of a copula method. In essence the copula method enables to write a joint probability distribution function with a tunable correlation between variables in a tractable form~\cite{Nelsen2006Introduction}. The copula method has been used in various disciplines such as finance~\cite{Embrechts2009Copulas}, engineering~\cite{Horvath2020CopulaBased}, biology~\cite{Ray2020CODC}, astronomy~\cite{Takeuchi2010Constructing, Takeuchi2020Constructing}, and time series analysis~\cite{Jo2019Analytically, Jo2019Copulabased}. However, due to the complexity of the formalism for the GFP, the analytical solution could be obtained only approximately, e.g., by assuming a locally tree structure for the underlying network~\cite{Jo2021Analytical}, implying that attributes of neighbors of the focal node are correlated with that of the focal node but not with each other. However, in reality, many social networks are highly clustered, i.e., abundant in triangles~\cite{Albert2002Statistical, Newman2003Why, Jo2018Stylized}. Thus attributes of neighbors of the focal node are expected to be correlated with each other too. This strongly calls for a more comprehensive analytical framework for the GFP in networks with triangles or clustered networks.

In this paper we propose such a framework by employing a vine copula method particularly for multiple correlated variables~\cite{Takeuchi2020Constructing}, by which in addition to the correlation between the focal node and its neighbors, one can also consider the correlation between attributes of neighbors of the focal node. We analytically and numerically find for the exponentially distributed attributes that the peer pressure of individuals with high (low) attributes is increased (reduced) by the connections between their neighbors. By our analytical approach we can get deeper insight into how the triangular structure of social networks affects individuals’ perception about their neighborhood, therefore better understand related phenomena observed in complex networks such as perception biases~\cite{Lee2019Homophily}. In addition, the copula method is expected to be useful for developing analytical approaches to various topics in complex systems.

We remark that for both FP and GFP, the degree or attribute of a node is compared to a set of degrees or attributes of its neighbors or a single value summarizing the set. The most common summarization is to take an average of degrees or attributes in the set, which is sensitive on few neighbors with a very high degree or attribute. The median of the set was also proposed as it is less sensitive on such neighbors than the average~\cite{Feld1991Why, Ugander2011Anatomy, Kooti2014Network, Momeni2016Qualities}. More recently, a novel summarization method in terms of the fraction of neighbors with a higher degree or attribute than the focal node has been suggested to study the effect of summarization methods on the FP~\cite{Lee2019Impact}. In our work, we focus on the most common method, namely, the mean-based method.

\section{Copula-based analysis}\label{sec:copula}

\begin{figure}[!t]
\includegraphics[width=0.6\columnwidth]{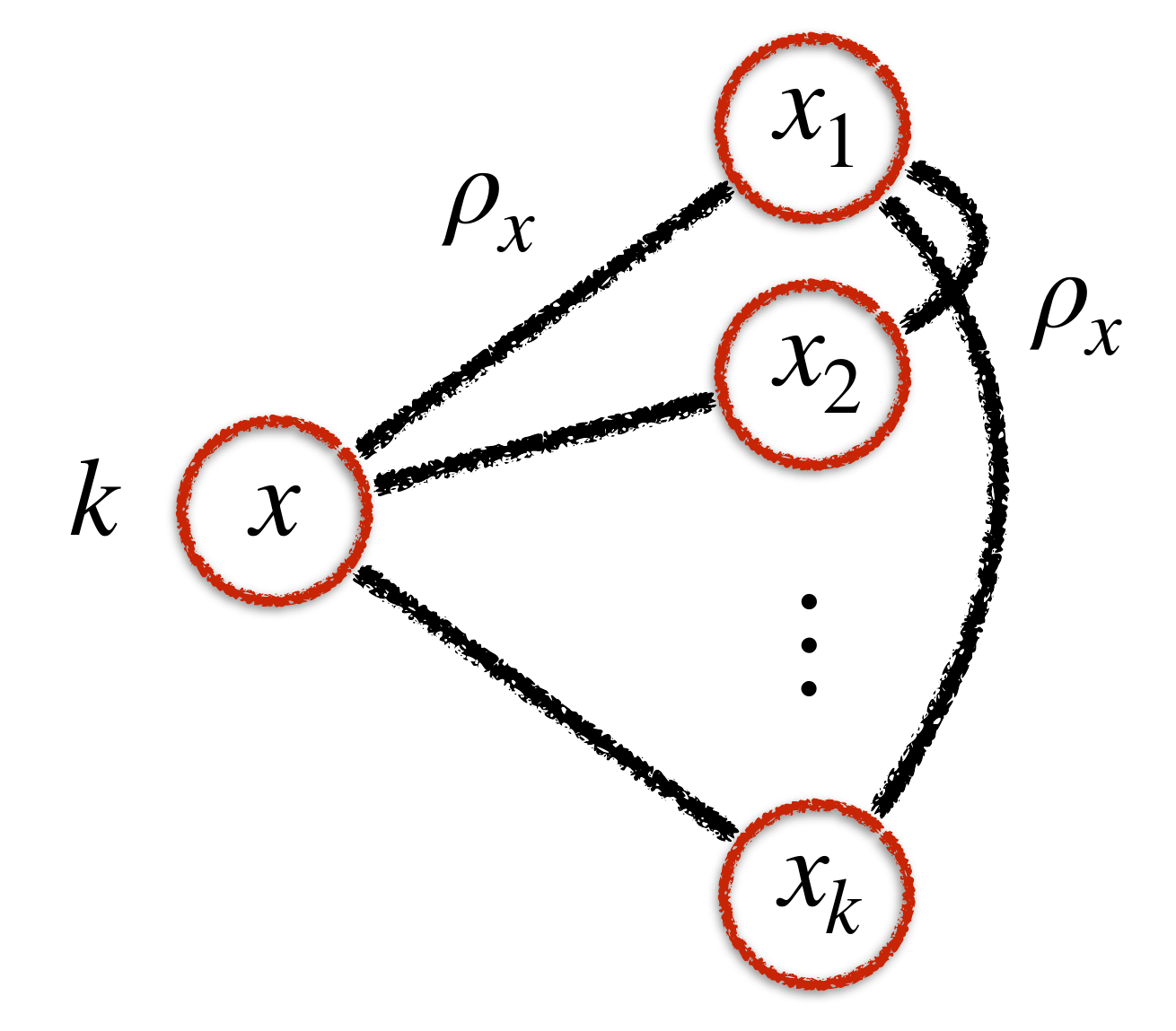}
\caption{Schematic diagram for the generalized friendship paradox in clustered networks. An ego with an attribute $x$ has $k$ neighbors whose attributes are $x_1,x_2,\ldots,x_k$, respectively. It is assumed to have the same correlation structure of attributes of neighboring nodes in every link, which is controlled by a single value of the Pearson correlation coefficient $\rho_x$ in Eq.~\eqref{eq:PCC}.}
\label{fig:schema}
\end{figure}

A network with $N$ nodes is considered. For non-topological attributes of such nodes, we consider an attribute distribution $P(x)$. For the sake of convenience, $x\geq 0$ is assumed. Yet values of $x<0$ can be considered too. At the individual level, the generalized friendship paradox (GFP) requires to compare an attribute of a focal node to a set of its neighbors' attributes or a single value representing the set. As mentioned, in this work we focus on the mean-based summarization method among others~\cite{Lee2019Impact}.

The mean-based GFP is said to hold for a node $i$ when the node's attribute is smaller than the average of its neighbors' attributes. Precisely, the mean-based GFP holds for the node $i$ when the condition below is satisfied~\cite{Eom2014Generalized}:
\begin{align}
  \frac{1}{k_i}\sum_{j\in\Lambda_i}x_j > x_i,
  \label{eq:mean_define_original}
\end{align}
where $\Lambda_i$ is the set of the node $i$'s neighbors, and $k_i\equiv |\Lambda_i|$ denotes the degree of the node $i$. Then the probability that Eq.~\eqref{eq:mean_define_original} is satisfied is called the mean-based peer pressure. One can write the mean-based peer pressure of the node having degree $k$ and attribute $x$ as
\begin{align}
  & h(k,x) \equiv \Pr\left(\frac{1}{k}\sum_{j=1}^k x_j>x\right) = \Bigg\langle \theta\left(\frac{1}{k}\sum_{j=1}^k x_j-x\right)\Bigg\rangle \nonumber \\
  & =\prod_{j=1}^k \int_0^\infty dx_j
  P(x_1,\ldots,x_k|x)\theta\left(\frac{1}{k}\sum_{j=1}^k x_j-x\right).
  \label{eq:mean_define}
\end{align}
Here $\theta(\cdot)$ denotes a Heaviside step function, and $\langle\cdot\rangle$ is the ensemble average over $\{x_1,\ldots,x_k\}$. $P(x_1,\ldots,x_k|x)$ denotes the conditional joint probability distribution function (PDF) of $k$ attributes of the focal node's neighbors, given the attribute $x$ of the focal node. See Fig.~\ref{fig:schema} for the schematic diagram. For modeling $P(x_1,\ldots,x_k|x)$, we adopt a C-vine copula to be introduced below.

We first introduce the simplest version of copula, i.e., a bivariate copula and then a vine copula for dealing with interdependence between multiple variables. The bivariate copula means a function $C$ that joins a bivariate cumulative distribution function (CDF), denoted by $F(x_1,x_2)$, to their one-dimensional marginal CDFs, denoted by $F_1(x_1)$ and $F_2(x_2)$~\cite{Nelsen2006Introduction, Cossette2013Multivariate}, such that
\begin{align}
    F(x_1,x_2) = C_{12}[F_1(x_1), F_2(x_2)].
\end{align}
The bivariate PDF of $x_1$ and $x_2$, denoted by $P(x_1,x_2)$, is then derived by 
\begin{align}
    &P(x_1,x_2) = \frac{\partial^2 C_{12}[F_1(x_1), F_2(x_2)]}{\partial x_1\partial x_2} \nonumber\\
    &\equiv c_{12}[F_1(x_1), F_2(x_2)] P_1(x_1)P_2(x_2),
\end{align}
where $P_1(x_1)$ and $P_2(x_2)$ denote PDFs. Denoting $F_1(x_1)$ and $F_2(x_2)$ by $u_1$ and $u_2$, the copula density $c_{12}$ is defined as
\begin{align}
    c_{12}(u_1,u_2) = \frac{\partial^2 C_{12}(u_1, u_2)}{\partial u_1\partial u_2},
\end{align}
which carries information on the correlation structure between $x_1$ and $x_2$.

We consider the case that a multivariate PDF with $d$ variables, i.e., $P(x_1,\ldots,x_d)$ for $d>2$, can be written in terms of a product of bivariate copulas. For this, we choose a C-vine copula among others~\cite{Takeuchi2020Constructing}:
\begin{widetext}
\begin{align}
    P(x_1,\ldots,x_d)=\prod_{j=1}^{d-1} \prod_{j'=j+1}^{d} c_{j,j'|1,\ldots,j-1}[F_{j|1,\ldots,j-1}(x_j|x_1,\ldots,x_{j-1}), F_{j'|1,\ldots,j-1}(x_{j'}|x_1,\ldots,x_{j-1})]\cdot \prod_{j=1}^d P_j(x_j),
    \label{eq:Cvine}
\end{align}
where $F_{j|1,\ldots,j-1}$ ($F_{j'|1,\ldots,j-1}$) denotes a conditional CDF of $x_j$ ($x_{j'}$) when $x_1,\ldots,x_{j-1}$ are given, and the conditional bivariate copula density $c_{j,j'|1,\ldots,j-1}$ carries information on the pairwise correlation between $x_j$ and $x_{j'}$ when $x_1,\ldots,x_{j-1}$ are given.

Now let us consider a focal node with an attribute of $x_0$ and $k$ neighbors, and the attributes of those neighbors are $\{x_1,\ldots,x_k\}$, respectively. For the analysis of the GFP for this node, we need to model a conditional joint PDF with $k$ variables $\{x_1,\ldots,x_k\}$ for a given $x_0$, i.e., $P(x_1,\ldots,x_k|x_0)$. Using a C-vine copula in Eq.~\eqref{eq:Cvine} one writes
\begin{align}
    P(x_1,\ldots,x_k|x_0)=\prod_{j=0}^{k-1} \prod_{j'=j+1}^{k} c_{j,j'|0,\ldots,j-1}[F_{j|0,\ldots,j-1}(x_j|x_0,\ldots,x_{j-1}), F_{j'|0,\ldots,j-1}(x_{j'}|x_0,\ldots,x_{j-1})]\cdot \prod_{j=1}^k P_j(x_j).
    \label{eq:jointPDF_define}
\end{align}
\end{widetext}
For the analytical tractability all $P_j$ for $j=1,\ldots,k$ are assumed to have the same form as 
\begin{align}
    P_j(x_j)=P(x_j).
    \label{eq:assume_P}
\end{align}
It is also assumed that $F_{j|0,\ldots,j-1}$ and $F_{j'|0,\ldots,j-1}$ have the same functional form, and that $x_j$ and $x_{j'}$ are not conditioned by $x_0,\ldots,x_{j-1}$, enabling us to write
\begin{align}
    &F_{j|0,\ldots,j-1}(x_j|x_0,\ldots,x_{j-1})= F(x_j),
    \label{eq:assume_F1}
    \\
    &F_{j'|0,\ldots,j-1}(x_{j'}|x_0,\ldots,x_{j-1})= F(x_{j'}),
    \label{eq:assume_F2}
\end{align}
where 
\begin{align}
    F(x_j)\equiv \int_0^{x_j}P(x)dx. 
\end{align}
Denoting $F(x_j)$ and $F(x_{j'})$ by $u_j$ and $u_{j'}$, we also assume that the copula density for a given $j$ has the same functional form for all $j'$:
\begin{align}
    c_{j,j'|0,\ldots,j-1}(u_j,u_{j'})= c_j(u_j,u_{j'}).
    \label{eq:cjj_assume}
\end{align}
Then $c_0(u_0,u_{j'})$ represents the correlation structure between the attribute of the focal node, $x_0$, and the attribute of its neighbor, $x_{j'}$, for $j'=1,\ldots,k$. On the other hand, $c_j(u_j,u_{j'})$ for $j\geq 1$ represents the correlation structure between attributes of the focal node's neighbors $j$ and $j'$ for $1\leq j<j'\leq k$. 

For modeling the pairwise copula density in Eq.~\eqref{eq:cjj_assume}, we adopt the Farlie-Gumbel-Morgenstern (FGM) copula density~\cite{Nelsen2006Introduction, Takeuchi2010Constructing, Cossette2013Multivariate}:
\begin{align}
    c_{_{\rm FGM}}(u,v)= 1+r(2u-1)(2v-1).
    \label{eq:FGM}
\end{align}
If $u$ and $v$ are, respectively, CDFs of variables $x$ and $x'$, $r\in [-1,1]$ controls the correlation strength between $x$ and $x'$. We relate $r$ to the Pearson correlation coefficient (PCC) between $x$ and $x'$. The PCC between $x$ and $x'$ is written as
\begin{align}
    \rho_{x}\equiv \frac{\langle xx'\rangle -\mu^2}{\sigma^2},
    \label{eq:PCC}
\end{align}
where 
\begin{align}
    \langle xx'\rangle\equiv \int_0^\infty dx \int_0^\infty dx' xx' P(x,x'),
\end{align}
and $\mu$ and $\sigma$ are the mean and standard deviation of $P(x)$, respectively. Using Eq.~\eqref{eq:FGM} one gets
\begin{align}
    \rho_{x}=\frac{r}{\sigma^2}\left\{\int_0^\infty dx x P(x)[2F(x)-1]\right\}^2\equiv Ar.
    \label{eq:rhoxx_r}
\end{align}
The upper bound of the value of $A$ is proven to be $1/3$ for any functional form of $P(x)$, which implies $|\rho_{x}|\le 1/3$~\cite{Schucany1978Correlation}.

Using the FGM copula, the case with $j=0$ in Eq.~\eqref{eq:cjj_assume} reads 
\begin{align}
    c_0(u_0,u_{j'})= 1+r(2u_0-1)(2u_{j'}-1).
    \label{eq:c0_FGM}
\end{align}
For the case with $j\geq 1$ in Eq.~\eqref{eq:cjj_assume}, we adopt the FGM copula density for the focal node's neighbors $j$ and $j'$: 
\begin{align}
    c_j(u_j,u_{j'})=\begin{cases} 1+r(2u_j-1)(2u_{j'}-1) & \textrm{for}\ a_{jj'}=1,\\
    1 & \textrm{otherwise},
    \end{cases}
    \label{eq:c1_FGM}
\end{align}
where $a_{jj'}=1$ indicates that $j$ and $j'$ are connected to each other. Note that we do not take into account any pairs of the focal node's neighbors that are not connected to each other. It is based on the assumption that the correlation is induced only by links between nodes.

Finally, using Eq.~\eqref{eq:jointPDF_define} with Eqs.~\eqref{eq:assume_P}--\eqref{eq:cjj_assume},~\eqref{eq:c0_FGM}, and~\eqref{eq:c1_FGM} we obtain
\begin{align}
    &P(x_1,\ldots,x_k|x_0)= \prod_{j=1}^k P(x_j) \cdot \prod_{j=1}^{k}[1+rG(x_0)G(x_j)]\nonumber\\ 
    &\times \prod_{\{jj'|a_{jj'}=1\}}[1+rG(x_j)G(x_{j'})],
    \label{eq:jointCondPDF}
\end{align}
where
\begin{align}
    G(x)\equiv 2F(x)-1.
\end{align}
Since it is not trivial to derive an exact form of Eq.~\eqref{eq:jointCondPDF}, by assuming $|r|\ll 1$, Eq.~\eqref{eq:jointCondPDF} is expanded up to the first order of $r$ :
\begin{align}
    &P(x_1,\ldots,x_k|x_0)\approx \prod_{j=1}^k P(x_j)\cdot \left[1+ rG(x_0)\sum_{j=1}^{k}G(x_j) \right.\nonumber\\
    &\left. + r \sum_{\{jj'|a_{jj'}=1\}}G(x_j)G(x_{j'}) + \mathcal{O}(r^2)\right].
    \label{eq:jointPDF}
\end{align}
We denote the size of the set $\{jj'|a_{jj'}=1\}$ as $n$, meaning the number of links between neighbors of the focal node. Then $0\leq n\leq k(k-1)/2$. We drop the subscript $0$ from $x_0$ hereafter.

We now take the Laplace transform of Eq.~\eqref{eq:mean_define} using Eq.~\eqref{eq:jointPDF} to get
\begin{align}
  & \tilde h(k,s)\approx \frac{1}{s}\left[1-\tilde P\left(\frac{s}{k}\right)^k\right] \nonumber\\ 
  & + rk\int_0^\infty dx_1 Q(x_1)
  \prod_{j=2}^k \int_0^\infty dx_j P(x_j) \int_0^{\bar x_k} dx e^{-sx} G(x) \nonumber \\
  & - rn \frac{1}{s} \tilde Q\left(\frac{s}{k}\right)^2 \tilde P\left(\frac{s}{k}\right)^{k-2}
  + \mathcal{O}(r^2),
  \label{eq:mean_laplace}
\end{align}
where 
\begin{align}
    Q(x)\equiv P(x)G(x),\ \bar x_k\equiv \frac{1}{k}\sum_{j=1}^k x_j,
\end{align}
and $\tilde P(s)$ and $\tilde Q(s)$ are the Laplace transforms of $P(x)$ and $Q(x)$, respectively. One can obtain the mean-based peer pressure $h(k,x)$ by means of the inverse Laplace transform of Eq.~\eqref{eq:mean_laplace}. We remark that the analytical result in Eq.~\eqref{eq:mean_laplace} was derived for the arbitrary functional form of $P(x)$ as well as for the arbitrary correlation coefficient $\rho_{x}$, the range of which is however limited by the shape of $P(x)$ [Eq.~\eqref{eq:rhoxx_r}]. We also mention that the approximations and assumptions made for the analysis may lead to the underestimation of the effects of correlations.

\section{Case study: Exponentially distributed attributes}\label{sec:case}

\subsection{Analytical solution}\label{subsec:anal}

For demonstrating a solvable case, we here consider an exponential distribution of the nodal attribute $x$, i.e.,
\begin{align}
  P(x)=\lambda e^{-\lambda x},
  \label{eq:expo_Px}
\end{align}
with $\langle x\rangle =1/\lambda$, leading to
\begin{align}
  Q(x)=\lambda e^{-\lambda x}-2\lambda e^{-2\lambda x}.
\end{align}
Note that $A=1/4$ from Eq.~\eqref{eq:rhoxx_r}, implying that 
\begin{align}
    \rho_x=\frac{r}{4}.
    \label{eq:rel_rhox_r}
\end{align}
Since 
\begin{align}
  \tilde P(s)=\frac{\lambda}{s+\lambda},\  \tilde Q(s)=\frac{-\lambda s}{(s+\lambda)(s+2\lambda)},
  \label{eq:expo_PsQs}
\end{align}
we obtain from Eq.~\eqref{eq:mean_laplace} 
\begin{align}
  &\tilde h(k,s) \approx \frac{1}{s}\left[1-\tilde P\left(\frac{s}{k}\right)^k\right] + rk \left[ -\frac{1}{s} \tilde P\left(\frac{s}{k}\right)^{k-1} \tilde Q\left(\frac{s}{k}\right)\right. \nonumber\\
  &\left. +\frac{2}{s+\lambda} \tilde P\left(\frac{s+\lambda}{k}\right)^{k-1} \tilde Q\left(\frac{s+\lambda}{k}\right) \right] \nonumber\\
  & +rn \left[-\frac{1}{s} \tilde P\left(\frac{s}{k}\right)^{k-2} \tilde Q\left(\frac{s}{k}\right)^2\right]
  +\mathcal{O}(r^2).
  \label{eq:mean_expo_laplace}
\end{align}

We then take the inverse Laplace transform of Eq.~\eqref{eq:mean_expo_laplace} to obtain an analytical solution of $h(k,x)$. Using Eq.~\eqref{eq:expo_PsQs} the inverse Laplace transform of the first term on the right hand side of Eq.~\eqref{eq:mean_expo_laplace} is given as
\begin{align}
    g(k,\lambda kx).
    \label{eq:h0}
\end{align}
Here $g(a,z)\equiv \Gamma(a,z)/\Gamma(a)$ denotes the regularized Gamma function, where $\Gamma(a)$ is the Gamma function and $\Gamma(a,z)=\int_z^\infty t^{a-1}e^{-t}dt$ is the upper incomplete Gamma function. 

The first term in the parentheses coupled to $rk$ on the right hand side of Eq.~\eqref{eq:mean_expo_laplace} is written as
\begin{align}
    -\frac{1}{s} \tilde P\left(\frac{s}{k}\right)^{k-1} \tilde Q\left(\frac{s}{k}\right) = \frac{1}{s+2\lambda k}\left(\frac{\lambda k}{s+\lambda k}\right)^k.
    \label{eq:app1}
\end{align}
To take the inverse Laplace transform of Eq.~\eqref{eq:app1}, we utilize the following result for the inverse Laplace transform~\cite{Jo2021Analytical}:
\begin{align}
    \mathcal{L}^{-1}\left\{
    -\frac{1}{s+b}\left(\frac{a}{s+a+b}\right)^k
    \right\}(x)
    =\left[g(k,ax)-1\right]e^{-bx}.
    \label{eq:invLaplace}
\end{align}
By plugging $a=-\lambda k$ and $b=2\lambda k$ into Eq.~\eqref{eq:invLaplace}, we get the inverse Laplace transform of Eq.~\eqref{eq:app1} as
\begin{align}
    (-1)^{k+1}\left[g(k,-\lambda k x)-1\right]e^{-2\lambda kx}.
    \label{appeq:1st}
\end{align}
The second term in the parentheses coupled to $rk$ on the right hand side of Eq.~\eqref{eq:mean_expo_laplace} is written as
\begin{align}
    & \frac{2}{s+\lambda} \tilde P\left(\frac{s+\lambda}{k}\right)^{k-1} \tilde Q\left(\frac{s+\lambda}{k}\right)\nonumber \\
    &= \frac{-2}{s+\lambda + 2\lambda k}\left(\frac{\lambda k}{s+\lambda+\lambda k}\right)^k.
    \label{eq:app2}
\end{align}
For the inverse Laplace transform of Eq.~\eqref{eq:app2} we plug $a=-\lambda k$ and $b=\lambda + 2\lambda k$ into Eq.~\eqref{eq:invLaplace} to get
\begin{align}
    2 (-1)^{k}\left[g(k,-\lambda k x)-1\right]e^{-(\lambda+2\lambda k)x}.
    \label{appeq:2nd}
\end{align}

The term coupled to $rn$ on the right hand side of Eq.~\eqref{eq:mean_expo_laplace} is explicitly written as
\begin{align}
    -\frac{1}{s} \tilde P\left(\frac{s}{k}\right)^{k-2} \tilde Q\left(\frac{s}{k}\right)^2 = -\frac{s}{(s+2\lambda k)^2}\left(\frac{\lambda k}{s+\lambda k}\right)^k,
    \label{eq:app3}
\end{align}
equivalently,
\begin{align}
    -\frac{1}{s+2\lambda k}\left(\frac{\lambda k}{s+\lambda k}\right)^k + 2\lambda k \frac{1}{(s+2\lambda k)^2}\left(\frac{\lambda k}{s+\lambda k}\right)^k.
    \label{eq:app3_expand}
\end{align}
The first term in Eq.~\eqref{eq:app3_expand} is the same as Eq.~\eqref{eq:app1} except for the sign, while for the second term we utilize Eq.~\eqref{eq:invLaplace} to derive
\begin{align}
    &\mathcal{L}^{-1}\left\{
    \frac{1}{(s+b)^2}\left(\frac{a}{s+a+b}\right)^k
    \right\}(x)\nonumber\\
    &=\left[-x g(k,ax)+x+\frac{k}{a}g(k+1,ax)-\frac{k}{a}\right]e^{-bx}.
    \label{eq:invLaplace2}
\end{align}
Then the inverse Laplace transform of the second term in Eq.~\eqref{eq:app3_expand} is obtained using Eq.~\eqref{eq:invLaplace2} by setting $a=-\lambda k$ and $b=2\lambda k$:
\begin{align}
    &2\lambda k (-1)^k \left[-x g(k,-\lambda k x)+x -\frac{1}{\lambda}g(k+1,-\lambda kx)+\frac{1}{\lambda}\right]\nonumber\\
    &\times e^{-2\lambda kx}.
    \label{appeq:4th}
\end{align}
Combining Eqs.~\eqref{eq:h0},~\eqref{appeq:1st},~\eqref{appeq:2nd}, and~\eqref{appeq:4th}, we finally get an analytical solution of $h(k,x)$ as follows:
\begin{align}
  h(k,x) \approx h_0(k,x) + r h_{\rm fn}(k,x) + rn h_{\rm nn}(k,x) +\mathcal{O}(r^2),
  \label{eq:hkx_mean_expo}
\end{align}
where
\begin{align}
    & h_0(k,x)\equiv g(k,\lambda kx), \label{eq:h0kx}\\
    & h_{\rm fn}(k,x)\equiv k(-1)^{k+1} e^{-2\lambda kx} (1-2e^{-\lambda x})\left[ g(k,-\lambda kx)-1 \right], \label{eq:h1kx}\\
    & h_{\rm nn}(k,x)\equiv (-1)^k e^{-2\lambda kx} \left\{(1-2\lambda kx)\left[g(k,-\lambda kx)-1\right]\right. \nonumber \\ 
    &\left.-2k \left[g(k+1,-\lambda kx)-1\right]\right\}. \label{eq:h2kx}
\end{align}
Here the subscript ``fn'' (``nn'') means between the focal node and its neighbor (between neighbors).

\begin{figure}[!t]
\includegraphics[width=\columnwidth]{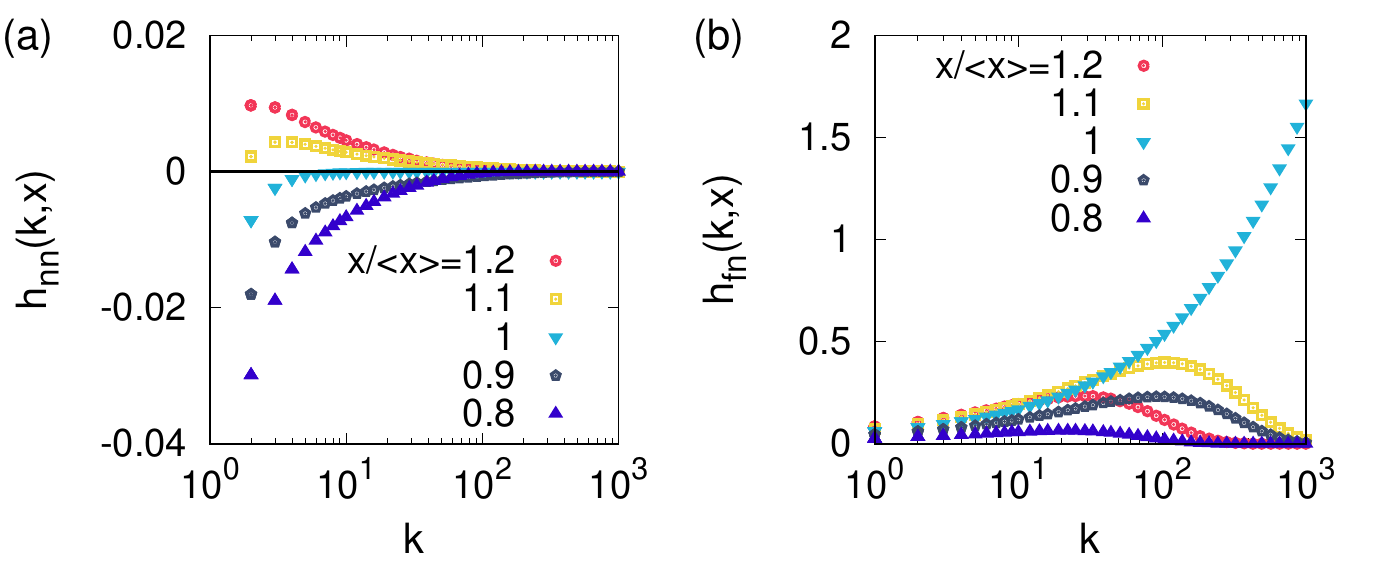}
\caption{Analytical results of $h_{\rm nn}(k,x)$ in Eq.~\eqref{eq:h2kx} (a) and
$h_{\rm fn}(k,x)$ in Eq.~\eqref{eq:h1kx} (b) as a function of $k$ for several values of $x$. In the panel (a) the results of $k=1$, leading to $n=0$, were omitted, and the black horizontal line denotes $0$ for guiding eyes.}
\label{fig:hkxn}
\end{figure}

\begin{figure*}[!t]
\includegraphics[width=\textwidth]{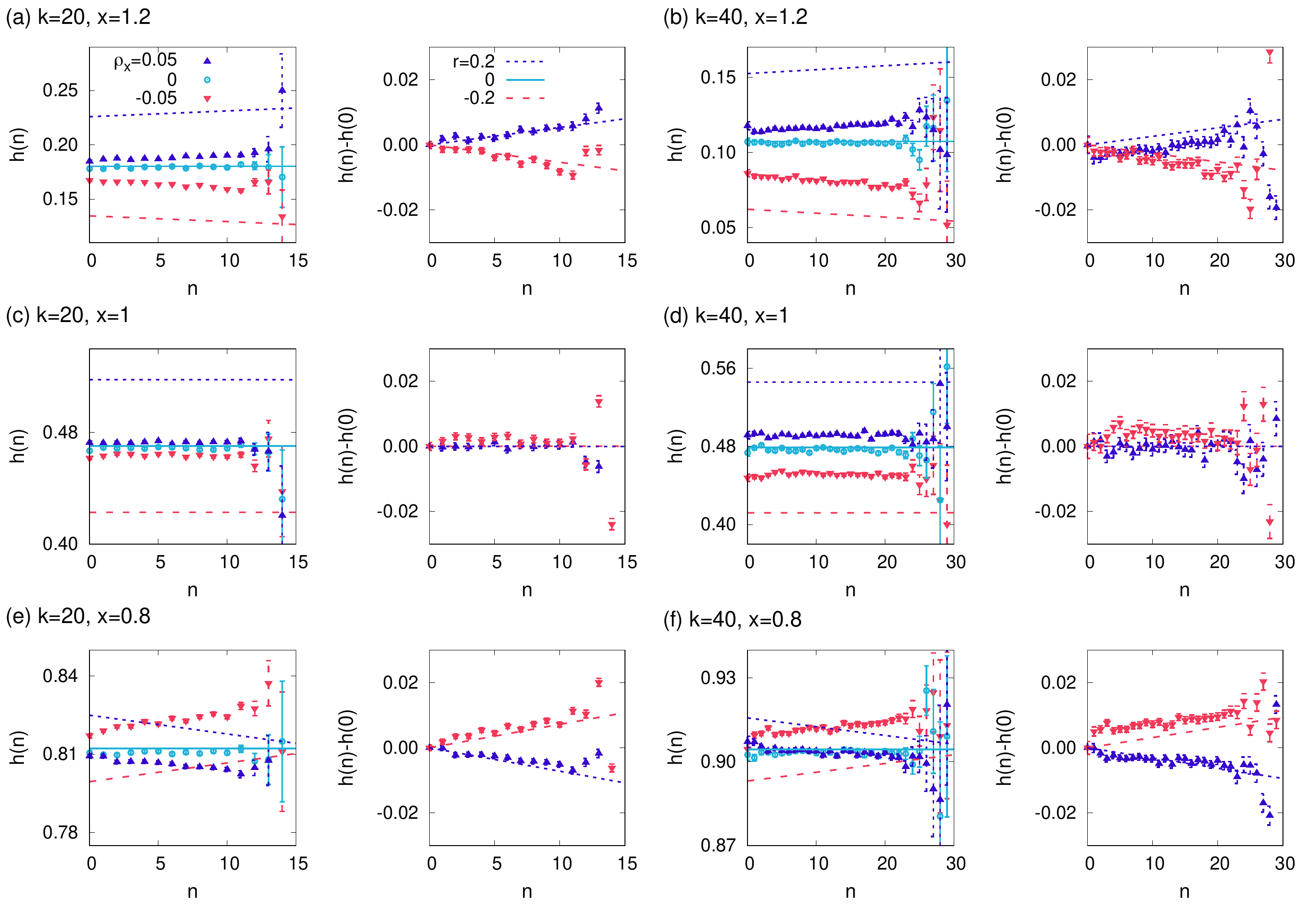}
\caption{Simulation results of the mean-based peer pressure $h(k,x)$ as a function of $n$ for values of $\rho_x=0.05$ (triangle), $0$ (circle), and $-0.05$ (inverse triangle), which we denote by $h(n)$ for clear presentation, and their differences from the results for $n=0$, denoted by $h(n)-h(0)$. $50$ clustered networks were generated with $N=5\times 10^4$ and average degree $50$, and $10^5$ different configurations of attributes for each network were randomly generated using $P(x)$ with $\lambda=1$ in Eq.~\eqref{eq:expo_Px} for each value of $\rho_x$. Error bars denote standard errors. For comparison, we plot the analytical solutions of $h(k,x)$ and $rnh_{\rm nn}(k,x)$ in Eq.~\eqref{eq:hkx_mean_expo} using $r=0.2$, $0$, and $-0.2$ [Eq.~\eqref{eq:rel_rhox_r}] as dotted, solid, and dashed lines, respectively. 
}
\label{fig:result}
\end{figure*}

By the above solution we can rigorously understand the effect of the attribute correlation between neighbors of a focal node on the mean-based peer pressure of the focal node. We first remark that the first and second terms on the right hand side of Eq.~\eqref{eq:hkx_mean_expo} are the same as the previous analytical solution obtained for a tree-like network with correlated attributes~\cite{Jo2021Analytical}. Therefore we focus on the third term on the right hand side of Eq.~\eqref{eq:hkx_mean_expo}, representing the effect of the number of links between neighbors of the focal node (i.e., $n$) on the mean-based peer pressure of the focal node. As shown in Fig.~\ref{fig:hkxn}(a), the value of $h_{\rm nn}(k,x)$ is positive for $x>\langle x\rangle$ and negligible for $x=\langle x\rangle$, while it is negative for $x< \langle x\rangle$. Therefore, from Eq.~\eqref{eq:hkx_mean_expo} the mean-based peer pressure can be either an increasing function of $n$ if $r>0$ and $x>\langle x\rangle$ or if $r<0$ and $x<\langle x\rangle$, a decreasing function of $n$ if $r>0$ and $x<\langle x\rangle$ or if $r<0$ and $x>\langle x\rangle$, or overall constant if $r=0$ and/or $x=\langle x\rangle$. These expectations will be tested against the simulation results in the next subsection. 

We remark that $h_{\rm nn}(k,x)$ turns out to vanish in the limit of large $k$ [see Fig.~\ref{fig:hkxn}(a)], which can be explained by the fact that if the focal node has considerably many neighbors, the average attribute of those neighbors eventually approaches the population mean of attributes $\langle x\rangle$, irrespective of the value of $\rho_x$ or $r$. Similarly, $h_{\rm fn}(k,x)$ vanishes for large $k$, except when $x=\langle x\rangle$, as shown in Fig.~\ref{fig:hkxn}(b). The divergent $h_{\rm fn}(k,x=\langle x\rangle)$ with an increasing $k$ is unphysical but expected to be canceled out by the higher-order terms of $r$. Therefore, in the limit of large $k$ the peer pressure of the focal node having a general $x$ is essentially determined by the result for the uncorrelated case, i.e., $h_0(k,x)$ in Eq.~\eqref{eq:h0kx}. The behavior of $h_0(k,x)$ for large $k$ was obtained in Eq.~(20) of Ref.~\cite{Jo2021Analytical}.

Let us now discuss implications of the results: In a network with positively correlated attributes ($\rho_x,r>0$), individuals having attributes higher (lower) than the population mean tend to have more (less) peer pressure when their neighbors are more connected to each other. In other words, the peer pressure of individuals with high (low) attributes is increased (reduced) by the connections between their neighbors. On the other hand, in a network with negatively correlated attributes ($\rho_x,r<0$) we observe the opposite tendency that individuals having attributes higher (lower) than the population mean tend to have less (more) peer pressure when their neighbors are more connected to each other and vice versa.

\subsection{Simulation results}\label{subsec:simul}

In order to test the analytical solution in Eq.~\eqref{eq:hkx_mean_expo}, we consider a random network model with tunable clustering, which is an extension of the configuration model independently suggested by Newman~\cite{Newman2009Random} and Miller~\cite{Miller2009Percolation}. For each node $i=1,\ldots,N$, the number of single links $s_i$ and the number of triangles $t_i$ are given, which are independently and randomly drawn from probability distributions $P(s)$ and $P(t)$, respectively. Then nodes are randomly chosen and connected to each other by single links or triangles incident to them. For the details of the model algorithm, see Refs.~\cite{Newman2009Random, Miller2009Percolation}. We use the \texttt{NetworkX} package in Python for implementation.  

We adopt geometric distributions for $s$ and $t$ as
\begin{align}
    &P(s)=(1-p)p^s,\\
    &P(t)=(1-q)q^t,
\end{align}
where $p,q\in [0,1)$ are parameters. The degree of the node $i$ is then obtained as $k_i=s_i+2t_i$, leading to the average degree $\langle k\rangle=\langle s\rangle + 2\langle t \rangle$ with $\langle s\rangle =p/(1-p)$ and $\langle t\rangle =q/(1-q)$.

For the simulation, we generate $50$ networks with $N=5\times 10^4$, $\langle s\rangle=30$, and $\langle t\rangle=10$, implying $\langle k\rangle=50$. Nodal attributes randomly drawn from the exponential distribution in Eq.~\eqref{eq:expo_Px} with $\lambda=1$ are assigned to nodes and then shuffled to get the Pearson correlation coefficient between neighboring attributes as some target value of $\rho_x$~\cite{Jo2021Analytical}. Here we use $\rho_x=-0.05$, $0$, and $0.05$. For each $\rho_x$, we randomly generate $10^5$ different configurations of attributes for each network. For each configuration, the mean-based peer pressure for each node $i$ is calculated as follows:
\begin{align}
h_{i}\equiv \theta\left(\frac{1}{k_i}\sum_{j\in\Lambda_i}x_j - x_i\right).
  \label{eq:hkx_mn_define}
\end{align}
Then we take the average of $h_i$ values for nodes with the same $k$ and $x$ as well as with the same $n$ over all configurations on all networks. 

As we focus on the effect of $n$ on $h(k,x)$, we plot the simulation results of $h(k,x)$ as a function of $n$ for combinations of parameter values of $k=20$, $40$ and $x=0.8$, $1$, $1.2$ as shown in Fig.~\ref{fig:result}. As expected, results for uncorrelated attributes ($\rho_x=0$) are in good agreement with the analytical solution for $r=0$. Numerical results of $h(k,x)$ for correlated attributes ($\rho_x\neq 0$) are qualitatively consistent with the analytical expectations for $r\neq 0$ in Eq.~\eqref{eq:hkx_mean_expo}. The deviations between simulation and analytical results for correlated cases are possibly due to approximations taken for deriving the analytical solution in Section~\ref{sec:copula}, such as the expansion of the equation up to the first-order terms of $r$ in Eq.~\eqref{eq:jointPDF}. Calculation of higher-order terms of $r$ is left for future work. In addition, considering the fact that the results for $n=0$ correspond to the case with a locally tree structure, the effects of triangles or $n$ on $h(k,x)$ can be delineated in terms of the difference between $h(k,x)$ for $n>0$ and that for $n=0$, which is denoted by $h(n)-h(0)$ in Fig.~\ref{fig:result}. We find that numerical differences of $h(k,x)$ for $n>0$ from that for $h=0$ are in good agreement with the corresponding analytical solution, i.e., $rnh_{\rm nn}(k,x)$ in Eq.~\eqref{eq:hkx_mean_expo}. Finally, we remark that large fluctuations in numerical results for large $n$ are due to the scarce samples for large $n$.

\section{Conclusion}\label{sec:concl}

The generalized friendship paradox (GFP) indicates that on average your neighbors have higher attributes than yours~\cite{Eom2014Generalized}. Despite successful demonstrations of the GFP by empirical analyses and numerical simulations, analytical, rigorous understanding of the GFP has been largely unexplored due to the lack of proper mathematical tools for modeling the correlation structure between attributes of neighboring nodes. Recently, an analytical solution for the probability of holding the GFP for an individual node in a network with correlated attributes was obtained using the copula method~\cite{Jo2021Analytical}. Here a locally tree structure of the underlying network was assumed for analytical tractability. However, the abundant triangles in most real social networks require a more general framework to incorporate the attribute correlation structure between neighbors of a focal individual in addition to the correlation between the focal individual and its neighbors. By employing a vine copula method, we have obtained the analytical solution for the GFP in clustered networks to find that the peer pressure of individuals with high (low) attributes is increased (reduced) by the connections between their neighbors. The analytical results were qualitatively supported by numerical simulations using the network models with tunable clustering. The deviation between analytical and numerical results could be accounted for by studying higher-order terms of the parameter controlling the correlation between attributes.

Our analytical approach can help us get deeper insight into how the triangular structure of social networks affects the GFP behavior, or more generally, individuals’ perception about their neighborhood. Hence it can lead to better understanding of various related phenomena in social networks such as perception biases~\cite{Lee2019Homophily}.

In our work we have studied the case with the exponential attribute distribution and the Farlie-Gumbel-Morgenstern copula for the attribute correlation, whereas the reality must be more complex than our assumptions. Indeed, the attribute distributions often show heavy tails~\cite{Clauset2009Powerlaw, Eom2014Generalized}. In addition, more realistic attribute correlation structures can be studied by employing other families of copulas~\cite{Nelsen2006Introduction}. Finally, we expect the vine copula method to be useful for modeling complex correlation structures between elements of complex systems.

\begin{acknowledgments}
H.-H.J. acknowledges financial support by the National Research Foundation of Korea (NRF) grant funded by the Korea government (MSIT) (No.~2022R1A2C1007358) and by the Catholic University of Korea, Research Fund, 2021.
Y.-H.E. acknowledges financial support by the National Research Foundation of Korea (NRF) grant funded by the Korea government (MSIT) (Grant No. 2020R1G1A1101950) and by Basic Science Research Program through the National Research Foundation of Korea (NRF) funded by the Ministry of Education (Grant No. 2018R1A6A1A06024977).
\end{acknowledgments}

%\bibliography{h2jo-papers}
%apsrev4-2.bst 2019-01-14 (MD) hand-edited version of apsrev4-1.bst
%Control: key (0)
%Control: author (8) initials jnrlst
%Control: editor formatted (1) identically to author
%Control: production of article title (0) allowed
%Control: page (0) single
%Control: year (1) truncated
%Control: production of eprint (0) enabled
%

\end{document}